\documentclass[conference]{IEEEtran}

\usepackage{amsmath}
\usepackage{amsmath,amsthm}
\usepackage{cite}   % Written by Donald Arseneau
\usepackage{graphicx}  % Written by David Carlisle and Sebastian Rahtz
\usepackage{amsmath, amsopn, psfrag, amsthm}   % From the American Mathematical Society
\usepackage{amssymb}
\usepackage{color}

\usepackage{algorithm}
\usepackage{amsmath}
\usepackage{algorithmic}
\usepackage[utf8]{inputenc}
\usepackage[english]{babel}
\newtheorem{theorem}{Theorem}

\newtheorem{proposition}{Proposition}
\usepackage{psfrag}
\usepackage{epsfig}
\graphicspath{{./img/}}
\DeclareGraphicsExtensions{.pdf,.jpeg,.png, .eps}
\usepackage{amssymb,amsmath,mathtools}

\def\bgama{{\boldsymbol{\Gamma}}}
\def\bPhi{{\boldsymbol{\Phi}}}

\def\bup{{\boldsymbol{\Upsilon}}}
\def\blam{{\boldsymbol{\Lambda}}}
\def\bz{{\boldsymbol{\zeta}}}
\def\bn{{\boldsymbol{\nu}}}
\def\bv{{\boldsymbol{\varrho}}}
\def\bk{{\boldsymbol{\kappa}}}

\usepackage{cite}   % Written by Donald Arseneau
\usepackage{graphicx}  % Written by David Carlisle and Sebastian Rahtz
\usepackage{amsmath, amsopn}   % From the American Mathematical Society
\usepackage{amssymb}
\usepackage{color}
\usepackage{psfrag}
\usepackage{epsfig}
\usepackage{stfloats}
\usepackage{lipsum}
\usepackage{multicol}
\graphicspath{{./img/}}
\DeclareGraphicsExtensions{.pdf,.jpeg,.png}
\begin{document}
\title{Cell-Free Massive MIMO with Limited Backhaul}
%\vspace{-0.001in}
%\linespread{.989}
\author{
\IEEEauthorblockN{Manijeh Bashar\IEEEauthorrefmark{1}, Kanapathippillai Cumanan\IEEEauthorrefmark{1}, Alister G. Burr\IEEEauthorrefmark{1}, Hien Quoc Ngo\IEEEauthorrefmark{2}, and Mérouane Debbah\IEEEauthorrefmark{3}}
\IEEEauthorblockA{\IEEEauthorrefmark{1}Department of Electronic Engineering, University of York, Heslington, York, UK, \IEEEauthorblockA{\IEEEauthorrefmark{2}School of Electronics, Electrical Engineering and Computer Science, Queen's University Belfast, Belfast, UK,\IEEEauthorblockA{\IEEEauthorrefmark{3}Large Networks and Systems Group, CentraleSupélec,
Université Paris-Saclay, France}}
Email:{ \{mb1465, kanapathippillai.cumanan, alister.burr\}@york.ac.uk}, hien.ngo@qub.ac.uk, m.debbah@centralesupelec.fr}
}
%
%\IEEEauthorblockA{\IEEEauthorrefmark{3}Mathematical and Algorithmic Sciences Lab, Huawei Technologies Co., Ltd., Boulogne-Billancourt 92100, France}
%Email: \{m.debbah@centralesupelec.fr, {merouane.debbah@huawei.com}\}}
\maketitle
\begin{abstract}
We consider a cell-free Massive multiple-input multiple-output (MIMO) system and investigate the system performance for the case when the quantized version of the estimated channel and the quantized received signal are available at the central processing unit (CPU), and the case when only the quantized version of the combined signal with maximum ratio combining (MRC) detector is available at the CPU.
Next, we study the max-min optimization problem, where the minimum user uplink rate is maximized with backhaul capacity constraints. To deal with the max-min non-convex problem, we propose to decompose the original problem into two sub-problems. 
Based on these sub-problems, we develop an iterative scheme which solves the original max-min user uplink rate. Moreover, we present a user assignment algorithm to further improve the performance of cell-free Massive MIMO with limited backhaul links.
\\\
\textcolor{white}{aa}{{\textbf{\textit{Keywords:}}} Cell-free Massive MIMO, geometric programming, generalized eigenvalue
problem, limited backhaul.}
\end{abstract}
\section{Introduction}
 \let\thefootnote\relax\footnotetext{The work of K. Cumanan and A. G. Burr was supported by H2020- MSCA-RISE-2015 under grant number 690750. In addition, the work on which this paper is based was carried out in collaboration with COST Action CA15104 (IRACON).}
Massive multiple-input multiple-output (MIMO) is a potential
technique to achieve high data rate \cite{5gdebbah,multidebbah,ourvtc18,ouricc1,ouriet_mic,slock_fromMU}.
Cell-free Massive MIMO is a promising technology for 5th Generation (5G) systems, where large number of distributed access points (APs) serve much smaller number of distributed users \cite{marzetta_free16}.
The effect of the limited capacity backhaul links from the APs to a central processing unit (CPU) has not however been addressed in the literature.
The assumption of infinite backhaul in \cite{marzetta_free16,ourvtc18,ouricc1,ourjournal2}
is not realistic in practice. The backhaul load is the main challenge in any distributed antenna systems \cite{backmacro}. First, we consider the case where all APs send back the quantized version of the  minimum mean square error (MMSE) estimate of the channel  from each user and the quantized version of the received signal to the CPU. 
We next study the case when each AP multiplies the received signal by the conjugate of the estimated channel from each user, and sends back a quantized version of this weighted signal to the CPU. We derive the total number of bits for both cases and show that given the same backhaul capacity for both cases, the relative performance of the aforementioned cases depends on the number of antennas at each AP, the total number of APs and the channel coherence time.
We next investigate an uplink max-min signal-to-interference plus noise ratio (SINR) problem with limited backhaul links, where to handle the non-convexity of the max-min SINR problem, we propose to decouple the original problem into two sub-problems, namely, receiver filter coefficient design, and power allocation. We next show that the receiver filter coefficient design problem may be solved through a generalized eigenvalue problem \cite{bookematrix} whereas the user power allocation problem is solved through a standard geometric programming (GP) \cite{gpboyd}, and present an iterative algorithm to alternately solve the max-min SINR problem. Finally an efficient user assignment algorithm  is presented, which results in significant performance improvement. 
%The rest of the paper is organized as follows. Section II describes
%the system model and Section III provides performance analysis. The proposed max-min SINR scheme is presented in Section IV and Section V investigates
%the proposed user assignment algorithm. Numerical results are presented in Section VI, and finally Section VII concludes the paper.
%\vspace{-.05in}
\section{SYSTEM MODEL}
We consider uplink transmission in a cell-free Massive MIMO system with $M$ APs and $K$ single-antenna users randomly distributed in a large area. Moreover, we assume each AP has $N$ antennas. The channel coefficient vector between the $k$th user and the $m$th AP, $\textbf{g}_{mk} \in \mathbb{C}^{N\times 1}$, is modeled as
$
\textbf{g}_{mk}=\sqrt{\beta_{mk}}\textbf{h}_{mk},
$
where $\beta_{mk}$ denotes the large-scale fading and $\textbf{h}_{mk}\sim  \mathcal{CN}(0,\bf{I}_N)$ represents the small-scale fading \cite{marzetta_free16}.
All pilot sequences transmitted by the $K$ users in the channel estimation phase are collected in a matrix $\bPhi \in \mathbb{C}^{\tau\times K}$, where $\tau$ is the length of the pilot sequence for each user and the $k$th column, $\pmb{\phi}_k$, represents the pilot sequence used for the $k${th} user. After performing a de-spreading operation, the MMSE estimate of the channel coefficient between the $k$th user and the $m$th AP is given by \cite{marzetta_free16}
\begin{IEEEeqnarray}{rCl}
\small
\!\!\!\!\!\!\hat{\textbf{g}}_{mk}\!=\!c_{mk}\!\left(\!\!\sqrt{\tau p_p}\textbf{g}_{mk}\!+\!\sqrt{\tau p_p}\sum_{k^\prime\ne k}^{K}\textbf{g}_{mk^\prime}\pmb{\phi}_k^H\pmb{\phi}_{k^\prime}\!+\!\textbf{W}_{p,m}\pmb{\phi}_k\!\right)\!,
\label{ghat}
\end{IEEEeqnarray}
where $\textbf{W}_{p,m}$ denotes the noise sequence at the $m$th antenna  whose elements are i.i.d. $\mathcal{CN}(0,1)$, $p_p$ represents the normalized signal-to-noise ratio (SNR) of each pilot sequence (which we define in Section VI), and $c_{mk}$ is given by
$
c _{mk}=\frac{\sqrt{\tau p_p}\beta_{mk}}{\tau p_p\sum_{k^\prime=1}^{K}\beta_{mk^\prime}|\pmb{\phi}_k^H{\pmb{\phi}}_{k^\prime}|^2+1}.
$
The estimated channels in (\ref{ghat}) are used by the APs to design the receiver coefficients and determine power allocations.
The transmitted signal from the $k$th user is represented by
$
x_k= \sqrt{q_k}s_k,
$
where $s_k$ ($\mathbb{E}\{|s_{k}|^2\} = 1$) and $q_k$ denotes the transmitted symbol and the transmit power from the \textit{k}th user, respectively. 
The $N\times 1$ received signal at the $m$th AP from all users is given by 
\begin{equation}
\textbf{y}_m= \sqrt{\rho}\sum_{k=1}^{K}\textbf{g}_{mk}\sqrt{q_k}s_k+\textbf{n}_m, 
\label{ym}
\end{equation}
where each element of $\textbf{n}_m \in \mathbb{C}^{N\times 1}$, $n_{n,m}\sim \mathcal{CN}(0,1)$ is the noise at the $m$th AP.
\\\\
{\textbf{Case 1.} \textit{Quantized Estimate of the Channel and Quantized Signal Available at the CPU}:}
The $m$th AP quantizes the terms $\hat{\textbf{g}}_{mk}$, $\forall k$, and $\textbf{y}_m$, and forwards the quantized CSI and the quantized signals in each symbol duration to the CPU. The quantized signal can be obtained as:
\begin{equation}
[\tilde{\textbf{y}}_{m}]_n= [\textbf{y}_{m}]_n+[\textbf{e}_{m}^y]_n=[\bz_{m}]_n+j[\bn_{m}]_n, ~\forall m ,n, 
\end{equation}
where $[\textbf{x}]_n$ represents the $n$th element of vector $x$, $[\textbf{e}_{m}^y]_n$ refers to the quantization error, and $[\bz_{m}]_n$ and $[\bn_{m}]_n$ are the real and imaginary parts of $[\textbf{y}_m]_n$, respectively. The analog-to-digital converter (ADC) quantizes the real and imaginary parts of $[\textbf{y}_m]_n$ with $\alpha$ bits each, which introduces quantization errors $[\textbf{e}_{m}^y]_n$ to the received signals \cite{Oppenheimsignal}. In addition, the ADC quantizes the MMSE estimate of CSI as:
\begin{equation}
[\tilde{\textbf{g}}_{mk}]_n\!=\![\hat{\textbf{g}}_{mk}]_n\!+\![\textbf{e}_{mk}^{\hat{g}}]_n\!=\![\bv_{mk}]_n+j[\bk_{mk}]_n,\forall k, n, 
\end{equation}
where $[\bv_{mk}]_n$ and $[\bk_{mk}]_n$ denote the real and imaginary parts of $[\tilde{\textbf{g}}_{mk}]_n$, respectively. The received signal for the $k$th user after using the low complexity maximum ratio combining (MRC) detector at
the CPU is given by
\begin{IEEEeqnarray}{rCl} \label{rkcase2}
r_k &=&\sum_{m=1}^{M}\tilde{\textbf{g}}_{mk}^H\tilde{\textbf{y}}_{m}=
\sum_{m=1}^M\left(\hat{\textbf{g}}_{mk}\nonumber +\textbf{e}_{mk}^{\hat{g}}\right)^H\left(\textbf{y}_m+\textbf{e}_{m}^y\right)\\\nonumber
 &\!=\!\!&\underbrace{\sqrt{\rho}\mathbb{E}\left\{\sum_{m=1}^M\hat{\textbf{g}}_{mk}^H
 {\textbf{g}}_{mk}\sqrt{q_k}\right\}}_{\text{DS}_k}s_k+
 \underbrace{\sum_{m=1}^{M}\hat{\textbf{g}}_{mk}^H\textbf{n}_m}_{\text{TN}_k}
 \nonumber\\
 &\!+\!&\underbrace{\sqrt{\rho}\!\left(\!\sum_{m=1}^M\!\hat{\textbf{g}}_{mk}^H\!{\textbf{g}}_{mk}\sqrt{q_k}\!-\!\mathbb{E}\!\left\{\!\sum_{m=1}^M\! \hat{\textbf{g}}_{mk}^H{\textbf{g}}_{mk}\sqrt{q_k}\!\right\}\right)}_{{\text{BU}_k}}\!s_k
 \nonumber\\
 &+&\sum_{k^{\prime}\neq k}^{K}\underbrace{\sqrt{\rho}\sum_{m=1}^M\hat{\textbf{g}}_{mk}^H
 {\textbf{g}}_{mk^\prime}\sqrt{q_{k^\prime}}}_{{\text{IUI}_{kk^\prime}}}s_{k^{\prime}}   \nonumber\\\nonumber
 &+&
 \sum_{k^{\prime}= 1}^{K}\underbrace{\sqrt{\rho}\sum_{m=1}^{M}(\textbf{e}_{mk}^{\hat{g}})^H{\textbf{g}}_{mk^\prime}\sqrt{q_{k^\prime}}
 s_{k^{\prime}}}_{\text{TQE}_{kk^\prime}}
+\underbrace{\sum_{m=1}^{M}(\textbf{e}_{mk}^{\hat{g}})^H\textbf{n}_m}_{\text{TQE}_k^g}\nonumber\\
&+&
 \underbrace{\sum_{m=1}^{M}\hat{\textbf{g}}_{mk}^H\textbf{e}_m^y}_{\text{TQE}_k^y}+
 \underbrace{\sum_{m=1}^{M}(\textbf{e}_{mk}^{\hat{g}})^H\textbf{e}_m^y}_{\text{TQE}_k^{gy}},
\end{IEEEeqnarray}
where $\text{DS}_k$ and $\text{BU}_k$ denote the desired signal (DS) and beamforming uncertainty (BU) for the $k$th user, respectively, and $\text{IUI}_k$ represents the inter-user-interference (IUI) caused by the $k^\prime$th user. In addition, $\text{TN}_k$ accounts for the total noise (TN) following the MRC detection, and finally the terms $\text{TQE}_k^\text{y}$, $\text{TQE}_k^\text{g}$, $\text{TQE}_k^{\text{gy}}$ and $\text{TQE}_{kk^\prime}$ refere to the total quantization error (TQE) at the $k$th user due to the quantization errors at the channel and signal.
\begin{proposition}
The terms $\text{DS}_k$, $\text{BU}_k$, $\text{IUI}_{kk^\prime}$, $\text{TQN}_{kk^\prime}$, $\text{TQN}_k^g$, $\text{TQN}_k^y$, $\text{TQN}_k^{gy}$ are mutually uncorrelated.
\end{proposition}
The proof uses the fact that the quantization error is signal
independent, uniformly distributed white noise \cite{Oppenheimsignal}, and is
omitted here due to space limitations. ~~~~~~~~~~~~~~~~~~~~~~~~~~~~~~~~~$\blacksquare$
\\\
Using Proposition 1 and the same scheme in \cite{marzetta_free16}, the SINR of the received signal in (\ref{rkcase2}) can be defined by considering the worst-case of the uncorrelated Gaussian noise as (\ref{sinrk_apx_quan}) (defined at the top of the next page).
\begin{figure*}[t!]
\begin{IEEEeqnarray}{rCl}
\hrulefill
\label{sinrk_apx_quan}
{\textrm{SINR}}_k^{\text{Case 1}}=
\dfrac{\left|\text{DS}_k\right|^2}
{\mathbb{E}\!\left\{\!\left|\text{BU}_k\right|^2\!\right\}\!+
\!\sum_{k^\prime\ne k}^K\mathbb{E}\!\left\{\left|\text{IUI}_{kk^\prime}\!\right|^2\!\right\}+
\!\mathbb{E}\left\{\left|\text{TQE}_k^\text{y}\!\right|^2\!\right\}+
\!\mathbb{E}\left\{\left|\text{TQE}_k^\text{g}\!\right|^2\!\right\}+
\!\mathbb{E}\left\{\left|\text{TQE}_k^\text{gy}\!\right|^2\!\right\}+
\sum_{k^\prime=1}^K
\!\mathbb{E}\left\{\left|\text{TQE}_{kk^\prime}\!\right|^2\!\right\}
\!}.
\end{IEEEeqnarray}
%\vspace{-.05cm}
\begin{IEEEeqnarray}{rCl}
\label{sinr_er}
\!\!\!\!\!\!\!\text{SINR}_k^{\text{Case 1}}\!\!=\!\dfrac{N^2q_k\left(\!\sum_{m=1}^{M}\!\gamma_{mk}\right)^2}{\!\!\!N^2\!\sum_{k^\prime\ne k}^K\!q_{k^\prime}\!\left(\!\!\sum_{m=1}^{M}\!\gamma_{mk}\!\dfrac{\!\beta_{mk^\prime}\!}{\beta_{mk}}\!\right)^2\!|\pmb{\phi}_k^H\pmb{\phi}_{k^\prime}|^2\!+\!N\sum_{m=1}^{M}\left(C_{\text{tot},m}\!+\!1\!\right)\gamma_{mk}\!\sum_{k^\prime=1}^{K}\!q_{k^\prime}\!\beta_{mk^\prime}\!+\!\dfrac{\!N\!}{\!\rho\!}\sum_{m=1}^{M}
\left(C_{\text{tot},m}\!+\!1\!\right)\gamma_{mk}}\!.
\end{IEEEeqnarray}
\end{figure*}
\begin{theorem}
\label{theorem_up_quan_case2}
Having the quantized CSI and the quantized signal at the CPU and employing MRC detection at the CPU, the SINR of the \textit{k}th user is given by (\ref{sinr_er}) (defined at the top of the next page).
\end{theorem}
{\textit{Proof:}}
The distribution of the errors, $\textbf{e}_{m}^y$ and $\textbf{e}_{mk}^{\hat{g}}$, are uniform over the range of the quantization errors \cite{Oppenheimsignal} and the elements of $[\textbf{e}_{m}^y]_n$ and $[\textbf{e}_{mk}^{\hat{g}}]_n$ are i.i.d. random variables with variance $\mathbb{E}\{|[\textbf{e}_{m}^y]_n|^2\}$ and  $\mathbb{E}\{|[\textbf{e}_{mk}^{\hat{g}}]_n|^2\}$, respectively. To calculate $\mathbb{E}\{|[\textbf{e}_{m}^y]_n|^2\}$, we use the following property of the quantization error \cite{Oppenheimsignal}
\begin{IEEEeqnarray}{rCl}
\mathbb{E}\left\{\left|\left[\textbf{e}_{m}^y\right]_n\right|^2\right\} = 2 \sigma_{\text{Re}([\textbf{e}_{m}^y]_n)}^2= 2 \sigma_{\text{Im}([\textbf{e}_{m}^y]_n)}^2= 2 \left(\dfrac{\Delta^2}{12}\right),
\label{deltay}
\end{IEEEeqnarray}
where $\sigma_{\text{Re}([\textbf{e}_{m}^y]_n)}^2$ and $\sigma_{\text{Im}([\textbf{e}_{m}^y]_n)}^2$ are the variance of real and imaginary parts of the quantization error, and $\Delta$ is given by
$
\Delta=\frac{\mathcal{R}(\text{Re}([\textbf{y}_{m}]_n))}{Q_m}=\frac{\mathcal{R}(\text{Im}([\textbf{y}_{m}]_n))}{Q_m},
$
where $Q_m$ refers to the quantization level, $\mathcal{R}$ is the operator range and $\mathcal{R}\left(\text{Re}([\textbf{y}_{m}]_n)\!\right)$ is obtained as \cite{Oppenheimsignal}
\begin{IEEEeqnarray}{rCl}
\small
\!\!\!\!\!\!\!\!\mathcal{R}\!\left(\text{Re}\left(\left[\textbf{y}_{m}\right]_n\right)\!\right)\!\!=\!\!w_{y}\!\sigma_{\text{Re}([\textbf{y}_{m}]_n}\!-\!\! w_{y}\!(-\sigma_{\text{Re}([\textbf{y}_{m}]_n}\!)\!\!=\!\!2w_{y}\sigma_{\text{Re}([\textbf{y}_{m}]_n}\!.\!\label{range21y}
\end{IEEEeqnarray}
Note that the same equality holds for the term $\text{Im}\left([\textbf{y}_{m}]_n\right)$. 
 Moreover, the same approach is used to find the relation between $\mathcal{R}(\text{Re}([\hat{\textbf{g}}_{mk}]_n))$ and the standard deviation of $\text{Re}([\hat{\textbf{g}}_{mk}]_n)$, $\sigma_{\text{Re}([\hat{\textbf{g}}_{mk}]_n)}$. 
The proper values of $w_{y}$ and $w_{g}$ are numerically obtained.
Using (\ref{range21y}) and (\ref{deltay}), the power of the quantization errors is given by
\begin{equation}
\sigma_{\left[\textbf{e}_{m}^y\right]_n}^2\!=\frac{\omega_y^2
\sigma_{\text{Re}\{[\textbf{y}_{m}]_n\}}^2}{3Q_m^2\!}=\frac{\omega_y^2\sigma_{\text{Im}\{[\textbf{y}_{m}]_n\!\}}^2}{3Q_m^2}.
\end{equation}
Next, the term $\sigma_{\textbf{y}_{m}}^2$ is obtained as
\begin{equation}
\sigma_{\textbf{y}_{m}}^2 = \mathbb{E}\left\{\textbf{y}_m^H\textbf{y}_m\right\}=
N\left(\rho\sum_{k^\prime=1}^{K}q_{k^\prime}\beta_{mk^\prime}+1\right).
\end{equation}
Therefore, we have $ \sigma_{[\textbf{y}_{m}]_n}^2=\rho\sum_{k^\prime=1}^{K}q_{k^\prime}\beta_{mk^\prime}+1$, which enables us to find the variance of the real and imaginary parts of $[\textbf{y}_m]_n$ as follows:
$\sigma_{\text{Re}\{[\textbf{y}_{m}]_n\}}^2=  \sigma_{\text{Im}\{[\textbf{y}_{m}]_n\}}^2=\frac{\rho\sum_{k^\prime=1}^{K}q_{k^\prime}\beta_{mk^\prime}+1}{2}.
$
Finally, using the same method as $[\textbf{e}_{m}^y]_n$ for $[\textbf{e}_{m}^{\hat{g}}]_n$, and the fact that $\sigma_{[\hat{\textbf{g}}_{mk}]_n}^2=\mathbb{E}\{ [\hat{\textbf{g}}_{mk}^H\hat{\textbf{g}}_{mk}]_n\}=\gamma_{mk}$, the variance of the real and imaginary parts of $[\hat{\textbf{g}}_{mk}]_n$ is obtained as follows:
$
\sigma_{\text{Re}\{[\hat{\textbf{g}}_{m}]_n\}}^2=\sigma_{\text{Im}\{[\hat{\textbf{g}}_{m}]_n\}}^2=\frac{\gamma_{mk}}{2}.
$
Hence, the power of quantization errors can be obtained as
%\vspace{-.09in}
\begin{subequations}
\begin{eqnarray}
&\mathbb{E}\left\{\left|\left[\textbf{e}_{m}^y\right]_n\right|^2\!\right\}=\dfrac{w_{y}^2}{3Q_m^2}(\rho\sum_{k^\prime=1}^{K}q_{k^\prime}\beta_{mk^\prime}+1), \label{quanerr31}\\
&~\mathbb{E}\left\{\left|[\textbf{e}_{mk}^g]_n\right|^2\!\right\}=\dfrac{w_{g}^2}{3Q_m^2}\gamma_{mk}\label{quanerr32}.~~~~~~~~~~~~~~~~~~~~~
\end{eqnarray}
\end{subequations}
Using (\ref{quanerr31}) and (\ref{quanerr32}) and the fact that quantization error is indepnedent with the input of the quantizer, after some mathematical manipulations (the proof is omitted here due to space limitations), we have:
\begin{IEEEeqnarray}{rCl}
&&\!\!\!\!\!\!\!\!\!\!\!\mathbb{E}\left\{\left|\text{TQE}_k^\text{y}\!\right|^2\!\right\}\!\!+\!
\!\mathbb{E}\left\{\left|\text{TQE}_k^\text{g}\!\right|^2\!\right\}\!\!+\!
\!\mathbb{E}\left\{\left|\text{TQE}_k^\text{gy}\!\right|^2\!\right\}\!+\!\!
\sum_{k^\prime=1}^K
\!\mathbb{E}\left\{\left|\text{TQE}_{kk^\prime}\!\right|^2\!\right\}\nonumber\\
&=&
N\sum_{m=1}^{M}C_{\text{tot},m}\gamma_{mk}\!\sum_{k^\prime=1}^{K}\!q_{k^\prime}\!\beta_{mk^\prime}+N\sum_{m=1}^{M}C_{\text{tot},m}\gamma_{mk},
\end{IEEEeqnarray}
where $C_{\text{tot},m}=\frac{\omega_y^2}{3Q_m^2}+\frac{\omega_g^2}{3Q_m^2}+\frac{\omega_y^2\omega_g^2}{9Q_m^4}$. The terms $\left|\text{DS}_k\right|^2$, $\mathbb{E}\!\left\{\!\left|\text{BU}_k\right|^2\!\right\}$, and $\mathbb{E}\!\left\{\left|\text{IUI}_{kk^\prime}\!\right|^2\!\right\}$ are obtained as the similar method in \cite{marzetta_free16}, which completes the proof of Theorem \ref{theorem_up_quan_case2}.~~~~$\blacksquare$
\\\\
{\textbf{Case 2.} \textit{Quantized Weighted Signal Available at the CPU:}}
The $m$th AP quantizes the terms $z_{m,k}= \hat{\textbf{g}}_{mk}^{H}\textbf{y}_m$, $\forall k$, and forwards the quantized signals in each symbol duration to the CPU as
$
z_{mk}= \hat{\textbf{g}}_{mk}^{H}\textbf{y}_m =a_{mk}+jb_{mk}, ~\forall k, 
$
where $a_{mk}$ and $b_{mk}$ represent the real and imaginary parts of $z_{mk}$. An ADC quantizes the real and imaginary parts of $z_{m,k}$ with $\alpha$ bits each, which introduces quantization errors to the received signals \cite{Oppenheimsignal}.
Let us consider the term $e_{mk}^z$ as the quantization error of the $m$th AP. Hence, the relation between $z_{mk}$ and its quantized version, $\tilde{z}_{mk}$, can be written as
$
\tilde{z}_{mk} = z_{mk}+e_{mk}^z. 
$
The aggregated received signal at the CPU can be written as
%\vspace{-.05in}
\begin{IEEEeqnarray}{rCl}
r_k &=&\sum_{m=1}^{M}\left( \underbrace{\hat{\textbf{g}}_{mk}^H\!\textbf{y}_{m}\!}_{z_{mk}} + e_{mk}^z \right).
 \label{rkcase3}
\end{IEEEeqnarray}
\begin{figure*}[t!]
\begin{IEEEeqnarray}{rCl}
\begin{split}
\small
\!\!\!\!\!\!\!\!\!\!\!\!\!\!\!\text{SINR}_k^{\!\text{Case 2}}\!\!=\!\!\dfrac{N^2q_k\left(\!\sum_{m=1}^{M}\!\gamma_{mk}\right)^2}{\!N^2\!\sum_{k^\prime\ne k}^K\!q_{k^\prime}\!\left(\!\sum_{m=1}^{M}\!\gamma_{mk}\!\dfrac{\!\beta_{mk^\prime}\!}{\beta_{mk}}\!\!\right)^2\!\!|\!\pmb{\phi}_k^H\pmb{\phi}_{k^\prime}|^2\!\!+\!\!N\sum_{m=1}^{M}\!\left(\!\frac{\omega_z^2\left(2\beta_{mk}-\gamma_{mk}\!\right)\!}{3Q_m^2}\!+\!\gamma_{mk}\!\!\right)\!\!\sum_{k^\prime=1}^{K}\!q_{k^\prime}\!\beta_{mk^\prime}\!+\!\dfrac{\!N}{\!\rho}\!\sum_{m=1}^{M}\!\!\left(\!\!\frac{\omega_z^2}{3Q_m^2}\!+\!1\!\right)\!\gamma_{mk}}\!.\!
\label{sinrcpustat}
\end{split}
\end{IEEEeqnarray}
%\vspace{-.22in}
\hrulefill
\end{figure*}
\vspace{-.05in}
\begin{theorem}
\label{theorem_up_quan_case3}
Having the quantized weighted signal at the CPU and employing MRC detection at the CPU, the achievable uplink rate of the \textit{k}th user in the cell-free Massive MIMO system is given by (\ref{sinrcpustat}) (defined at the top of this page).
\end{theorem}
{\textit{Proof:}}  This can be derived by following the same approach for uplink transmission in Theorem 3.~~~~~~~~~~~~~~~~~~~~~~~~~~~~~~~$\blacksquare$
\subsection{Performance of Different Cases of Uplink Transmission}
Let us assume the length of frame (which represents the length of the uplink data) is $
\tau_f = \tau_c - \tau,
$
where $\tau_c$ denotes the number of samples for each coherence interval and $\tau$ represents the length of pilot sequence. Defining the number of the quantization levels as $Q_{m,i}=2^{\alpha_i}$, for $i = 1,2$, corresponding to Cases 1 and 2, for Case 1, the required number of bits for each AP during each coherence interval is $2\alpha_{1}\times(NK+N\tau_f)$ whereas Case 2 requires $2\alpha_{2}\times(K\tau_f)$ bits for each AP during each coherence interval. Hence, the total backhaul capacity required between the $m$th AP and the CPU for all schemes is defined as
\vspace{-.01in}
\begin{equation}
\small
C_{m}\!=\!\left\{
\begin{array}{rl} 
\dfrac{2\left(NK+N\tau_f\right)\log_2 Q_{m,1}}{T_c}, &\text{Case 1},\\\
\dfrac{2\left(K\tau_f\right)\log_2 Q_{m,2}}{T_c},~~~~ &\text{Case 2},
\end{array} \right.
\label{fr}
\end{equation}
where $T_c$ (in sec.) refers to coherence time. In the following, we present a comparison between three cases of uplink transmission. To make a fair comparison between Case 1 and Case 2, we use the same total number of backhaul bits for both cases, that is:
$
2(NK+N\tau_f)\log_2 Q_{m,1} = 2(K\tau_f)\log_2 Q_{m,2}.
$
In numerical results, we show that for the same backhaul capacity, the performances of Case 1 and Case 2 depend on the values of $N$, $K$ and $\tau_f$. In this work, we study the max-min SINR problem for Case 2 of uplink transmission in cell-free Massive MIMO system. 
%\begin{figure}[t!]
%	\center
%	\includegraphics[width=60mm]{new2_n4_k20_92_hien.eps}
%	\vspace{-0.14in}
%	\caption{The average sum rate for different uplink transmission schemes, with $N=4$, $K=20$, $D=1$ km, $\tau=20$, $\tau_c=200$, $\alpha_2=9$, $\alpha_3=2$, $\omega_g=3$, $\omega_y=80$ and $\omega_z=3$.}
%	\label{bit92}
%\end{figure}
%\begin{figure}[t!]
%	\center
%	\includegraphics[width=60mm]{new2_n20_k40_95_hien.eps}
%	\vspace{-0.17in}
%	\caption{The average sum rate for different uplink transmission schemes, with $N=20$, $K=40$, $D=1$ km, $\tau=40$, $\tau_c=200$, $\alpha_2=8$, $\alpha_3=5$, $\omega_g=3.5$, $\omega_y=70$ and $\omega_z=3$.}
%	\label{bit95}
%\end{figure} 
\section{Performance Analysis}
In this section, we derive the achievable rate for the considered system model in the previous section by following a similar approach to that in \cite{marzetta_free16}. In deriving the achievable rates of each user, it is assumed that the CPU exploits only the knowledge of channel statistics between the users and APs in detecting data from the received signal in (\ref{rk2}). The aggregated received signal at the CPU can be written as
\vspace{-.1in}
\begin{IEEEeqnarray}{rCl}
\!r_k\!&\!=\!\!&\sum_{m=1}^{M}\!u_{mk}\!\left(\!\underbrace{\!\hat{\textbf{g}}_{mk}^H\!\textbf{y}_{m}\!}_{z_{mk}}\!+\!e_{mk}^z\!\right)
\!=\!\nonumber\\\vspace{-.11in}
 &\!=\!\!&\underbrace{\sqrt{\rho}\mathbb{E}\left\{\sum_{m=1}^Mu_{mk}
 \hat{\textbf{g}}_{mk}^H{\textbf{g}}_{mk}\sqrt{q_k}\right\}}_{\text{DS}_k}s_k+
 \underbrace{\sum_{m=1}^{M}u_{mk}\hat{\textbf{g}}_{mk}^H\textbf{n}_m}_{\text{TN}_k}
 \nonumber\\\vspace{-.11in}
 &\!+\!&\underbrace{\sqrt{\rho}\!\left(\!\sum_{m=1}^M\!u_{mk}\hat{\textbf{g}}_{mk}^H\!{\textbf{g}}_{mk}\sqrt{q_k}\!-\!\mathbb{E}\!\left\{\!\sum_{m=1}^M\! u_{mk}\hat{\textbf{g}}_{mk}^H{\textbf{g}}_{mk}\sqrt{q_k}\!\right\}\right)}_{{\text{BU}_k}}\!s_k
 \nonumber\\\vspace{-.09in}
 &+&\sum_{k^{\prime}\neq k}^{K}\underbrace{\sqrt{\rho}\sum_{m=1}^Mu_{mk}\hat{\textbf{g}}_{mk}^H
 {\textbf{g}}_{mk^\prime}\sqrt{q_{k^\prime}}}_{{\text{IUI}_{kk^\prime}}}s_{k^{\prime}} + \underbrace{\sum_{m=1}^{M}u_{mk}e_{mk}^z}_{\text{TQE}_k},
 \label{rk2}
\end{IEEEeqnarray}
where by collecting all the coefficients $u_{mk}, ~\forall ~m$, corresponding to the \textit{k}th user, we define $\textbf{u}_k = [u_{1k}, u_{2k},\cdots, u_{Mk}]^T$ and without loss of generality, it is assumed that $|| \textbf{u}_k||=1$. The optimal solution of $\mathbf{u}_{k}, q_k,~\forall~k$ for the considered max-min SINR approach is investigated in Section IV. The corresponding SINR of the received signal in (\ref{rk2}) can be defined by considering the worst-case of the uncorrelated Gaussian noise as follows:
\vspace{-.06in}
\begin{IEEEeqnarray}{rCl}
\vspace{-.05in}
\label{sinrdef11}
\small
&&{\textrm{SINR}}_k= \\
&&\dfrac{\left|\text{DS}_k\right|^2}{\mathbb{E}\!\left\{\!\left|\text{BU}_k\right|^2\!\right\}\!+\!\sum_{k^\prime\ne k}^K\mathbb{E}\!\left\{\left|\text{IUI}_{kk^\prime}\!\right|^2\!\right\}\!+\!\mathbb{E}\left\{\left|\text{TN}_k\!\right|^2\!\right\}\!+\!\mathbb{E}\left\{\left|\text{TQE}_k\!\right|^2\!\right\}\!}.\nonumber
\end{IEEEeqnarray}
Based on the SINR definition in (\ref{sinrdef11}), the achievable uplink rate of the \textit{k}th user is defined in the following theorem.
\begin{figure*}
\begin{IEEEeqnarray}{rCl}
\small
R_k = \log_2\left( 1+\dfrac{\textbf{u}_k^H\left(N^2q_k\bgama_k\bgama_k^H\right)\textbf{u}_k}{\textbf{u}_k^H\left(N^2\sum_{k^\prime\ne k}^Kq_{k^\prime}|\pmb{\phi}_k^H\pmb{\phi}_{k^\prime}|^2\blam_{k k^\prime}\blam_{k k^\prime}^H+N\sum_{k^\prime=1}^{K}q_{k^\prime}\bup_{kk^\prime}+\dfrac{N}{\rho}\textbf{R}_{k}\right )\textbf{u}_k}\right).
\label{sinr1}
\end{IEEEeqnarray}
\vspace{-.66cm}
\hrulefill
\end{figure*}
\begin{theorem}
\label{theorem_up_quan_u}
Employing MRC weighting at APs, the achievable uplink rate of the \textit{k}th user in the Cell-free Massive MIMO system is given by (\ref{sinr1}) (defined at the top of this page). 
\end{theorem}
Note that $\textbf{u}_k=[u_{1k}, u_{2k}, \cdots, u_{Mk}]^T$, and the following equations hold:
$\bgama_k=[\gamma_{1k}, \gamma_{2k}, \cdots, \gamma_{Mk}]^T$,
$\bup_{kk^\prime}\!=\text{diag}\Big[\!\beta_{1k^\prime}(\frac{\omega_z^2(2\beta_{1k}\!-\!\gamma_{1k}\!)\!}{3Q_1^2}\!+\!\gamma_{1k}),\!\cdots$,
$\beta_{Mk^\prime}(\frac{\omega_z^2(2\beta_{Mk}-\gamma_{Mk}\!)\!}{3Q_M^2}\!+\!\gamma_{Mk})\!\Big] \nonumber$,
$\blam_{k k^\prime}=[\dfrac{\gamma_{1k}\beta_{1k^\prime}}{\beta_{1k}}, \dfrac{\gamma_{2k}\beta_{2k^\prime}}{\beta_{2k}}, \cdots, \dfrac{\gamma_{Mk}\beta_{Mk^\prime}}{\beta_{Mk}}]^T$ and
$\textbf{R}_{k} = \text{diag}\left[(\frac{\omega_z^2}{3Q_1^2}+1)\gamma_{1k}, \cdots, (\frac{\omega_z^2}{3Q_M^2}+1)\gamma_{Mk}\right]$.
\\\\
{\textit{Proof:}} Please refer to Appendix A.~~~~~~~~~~~~~~~~~~~~~~~~~~~
~~~~~~$\blacksquare$
\section{Proposed Max-Min SINR Scheme}
\label{secop}
In this section, we formulate the max-min SINR problem in cell-free massive MIMO, where the minimum uplink rates of all users is maximized while satisfying the transmit power constraint at each user and the backhaul capacity constraint as follows:
\vspace{-.08in}
\begin{subequations}
\label{p1} 
\begin{align}P_1:~~
\label{p1_1}\max_{q_k, \textbf{u}_k,Q_m}~~~~ &\min_{k=1,\cdots,K}\quad  R_k^{\text{UP}} ,\\
\label{p1_2}\text{s.t.}\quad &||\textbf{u}_k||=1, ~ \forall k,~~~\\
\label{p1_3}&0 \le q_k \le p_{\text{max}}^{(k)},  ~~ \forall k,\\
\label{p1_4}&C_m \le C_{\text{bh}},  ~~ \forall m,
\end{align}
\end{subequations}
where $p_{\text{max}}^{(k)}$ and $C_{\text{bh}}$ refer to the maximum transmit power available at user \textit{k} and the capacity of the backhaul link between the $m$th AP and the CPU, respectively.
It is obvious that the achievable user rates monotonically increase with the capacity of the backhaul link between the $m$th AP and the CPU. Hence, the optimal solution is achieved when $C_m = C_{\text{bh}},~\forall m$, which leads to fixed values for the number of quantization levels, $Q_m, \forall m$. As a result, the max-min SINR problem can be re-formulated as follows:
\vspace{-.15cm}
\begin{subequations}
\label{p2} 
\begin{align}P_2:~~~~~
\label{p2_1}\max_{q_k, \textbf{u}_k}~~~~ &\min_{k=1,\cdots,K}\quad  R_k^{\text{UP}} ,\\
\label{p2_2}~~~~~~~~~~~~\text{s.t. }\quad &||\textbf{u}_k||=1, ~ \forall k, ~0 \le q_k \le p_{\text{max}}^{(k)},  ~~ \forall k.
\end{align}
\end{subequations}
Problem $P_{2}$ is not jointly convex in terms of $\mathbf{u}_{k}$ and power allocation $q_{k},~\forall k$. Therefore, it cannot be directly solved through existing convex optimization software. To tackle this non-convexity issue, we decouple Problem $P_{2}$ into two sub-problems: receiver coefficient design (i.e. $\mathbf{u}_{k}$) and the power allocation problem. The optimal solution for Problem $P_{2}$, is obtained through alternately solving these sub-problems, as explained in the following subsections.
\subsection{Receiver Filter Coefficient Design}
In this subsection, the problem of designing the receiver coefficient is considered. We solve the max-min SINR problem for a given set of allocated powers at all users, $q_k, \forall k$, and fixed values for the number of quantization levels, $Q_m,~\forall m$. These coefficients (i.e., $\mathbf{u}_{k}$, $\forall~k$) are obtained by interdependently maximizing the uplink SINR of each user. Therefore, the optimal receiver filter coefficients can be determined by solving the following optimization problem:

\begin{subequations}
\vspace{-.1in}
\small
\label{p3} 
\begin{align}
&P_3:~ \max_{\mathbf{u}_k}&\nonumber\\
\label{p3_1}&\!\dfrac{\!N^2\textbf{u}_k^H\left(q_k\bgama_k\bgama_k^H\right)\textbf{u}_k\!}{\!\!\textbf{u}_k^H\!\left(\!N^2\!\sum_{k^\prime\ne k}^K\!q_{k^\prime}\!|\pmb{\phi}_k^H\pmb{\phi}_{k^\prime}|^2\!\blam_{k k^\prime}\!\blam_{k k^\prime}^H\!+\!N\!\sum_{k^\prime=1}^{K}q_{k^\prime}\bup_{kk^\prime}\!+\!\dfrac{N}{\rho}\!\textbf{R}_{k}\!\right)\!\textbf{u}_k\!}&\\
\label{p3_2}&\text{s.t. }||\textbf{u}_k||=1,~~~\forall k.&
\end{align}
\end{subequations}
Problem $P_{3}$ is a generalized eigenvalue problem \cite{cuma_sinr_spl10}, where the optimal solutions can be obtained by determining the generalized eigenvalue \cite{bookematrix} of the matrix pair $\mathbf{A}_{k} = N^2q_k\bgama_k\bgama_k^H$ and $\mathbf{B}_{k}\!=N^2\sum_{k^\prime\ne k}^K\!q_{k^\prime}\!|\pmb{\phi}_k^H\pmb{\phi}_{k^\prime}|^2\!\blam_{k k^\prime}\!\blam_{k k^\prime}^H\!+\!N\sum_{k^\prime=1}^{K}q_{k^\prime}\bup_{kk^\prime}\!+\!\frac{N}{\rho}\!\textbf{R}_{k}$ corresponding to the maximum generalized eigenvalue.
\subsection{Power Allocation} 
In this subsection, we solve the power allocation problem for a given set of fixed receiver filter coefficients, $\mathbf{u}_{k}$, $\forall~k$, and fixed values of quantization levels, $Q_m,~\forall m$. The optimal transmit power can be determined by solving the following max-min problem:
\begin{subequations}
\label{p4} 
\begin{align}P_4:
\label{p4_1}\max_{q_k}~~ \min_{k=1,\cdots,K}\quad & \text{SINR}_k^{\text{UP}} ,\\
\label{p4_2}\text{s.t. }\quad &0 \le q_k \le p_{\max}^{(k)}.~~~~~~
\end{align}
\end{subequations}
Without loss of generality, Problem $P_3$ can be rewritten by introducing a new slack variable as
\begin{subequations}
\label{p5} 
\begin{align}P_5:
\label{p5_1}\max_{t,q_k}\quad & t ,\quad &\\
\label{p5_2}\text{s.t. }\quad &0\le q_k \le p_{\max}^{(k)},~ \forall k,~\text{SINR}_{k}^{\text{UP}} \ge t, \forall k.
\end{align}
\end{subequations}
\begin{proposition}
Problem $P_{5}$ can be formulated into a GP. 
\end{proposition}
\vspace{-.1cm}
Therefore, Problem $P_5$ is efficiently solved through existing convex optimization software. Based on these two sub-problems, an iterative algorithm has been developed as summarized in Algorithm \ref{al1}. Note that $\epsilon$ in Step 2 of Algorithm \ref{al1} refers to a small predetermined value. In addition,
numerical results will be presented in
Section VI to validate the convergence of the proposed algorithm.
\begin{algorithm}[t]
\caption{Proposed algorithm to solve Problem $P_2$}

\hrulefill

\textbf{1.} Initialize $\textbf{q}^{(0)}=[q_1^{(0)},q_2^{(0)},\cdots,q_K^{(0)}]$, $i=1$

\textbf{2.} Repeat steps 3-5 until $\text{SINR}_{k}^{\text{UP},(i+1)}-\text{SINR}_{k}^{\text{UP},(i)}\le \epsilon, \forall k$

\textbf{3.} $i=i+1$

\textbf{4.} Set $\textbf{q}^{(i)}=\textbf{q}^{(i-1)}$ and determine the optimal receiver coefficients $\textbf{U}^{(i)}=[\textbf{u}^{(i)}_1,\textbf{u}^{(i)}_2,\cdots,\textbf{u}^{(i)}_K]$ through solving the generalized eigenvalue Problem $P_3$ in (\ref{p3})

\textbf{5.} Compute $\textbf{q}^{(i+1)}$ through solving Problem $P_5$ in (\ref{p5})

\hrulefill
\label{al1}
\end{algorithm}
\section{User Assignment}
The total backhaul capacity required between the $m$th AP and the CPU increases linearly with the total number of users served by the $m$th AP, which motivates the need to pick a proper set of active users for each AP. Using (\ref{fr}), we have

\begin{equation}
\log_2 Q_m\times K_{m}\le \dfrac{C_{\text{bh}}T_s \tau_c}{2K\tau_f},
\label{qk}
\end{equation}
where $K_m$ denotes the size of the set of active users for the $m$th AP.
From (\ref{qk}), it can be seen that decreasing the size of the set of active users 
allows for a larger number of quantization levels. Motivated by this fact, and to exploit the capacity of backhaul links more efficient, we investigate all possible combinations of $\log_2 Q_m$ and $K_{m}$. First, for a fixed value of $\log_2 Q_m$, we find an upper bound on the size of the set of active users for each AP. In the next step, we propose for all APs that the users are sorted according to $\beta_{mk},~\forall k$, and find the $K_{m}$ users which have the highest values of $\beta_{mk}$ among all users. If a user is not selected by any AP, we propose to find the AP which has the best link to this user. Then, we add the user to the set of active users for this user and drop the user which has the lowest $\beta_{mk},~\forall k$, among the set of active users for that AP which has links to other APs as well. We next solve the original max-min SINR problem with $\tilde{\gamma}_{mk}\gets \gamma_{mk}$.
\section{Numerical Results and Discussion}
A cell-free Massive MIMO system with $M$ APs and $K$ single antenna users is considered in a $D \times D$ simulation area, where both APs and users are uniformly distributed
at random. In the simulation, an uncorrelated shadowing model and a three-slope model for the path loss similar to \cite{marzetta_free16} are considered. Moreover, for the noise power, we use similar parameters as in \cite{marzetta_free16}. It is assumed that that $\bar{p}_p$ and $\bar{\rho}$ denote the power of the pilot sequence and the uplink data, respectively, where $p_p=\frac{\bar{p}_p}{p_n}$ and $\rho=\frac{\bar{\rho}}{p_n}$. In simulations, we set $\bar{p}_p=200$ mW and $\bar{\rho}=200$ mW. In addition, simulation results show $\omega_z=15$ is a proper value. Unless otherwise indicated, we set $\omega_z=15$.
\begin{figure}[t!]
	\center
	\includegraphics[width=80mm]{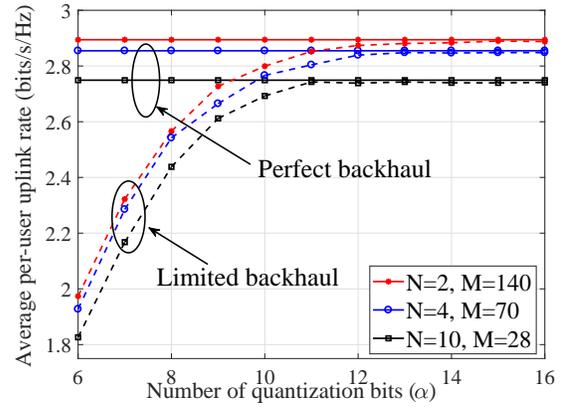}
	\vspace{-.09in}
	\caption{Average per-user uplink rate versus the number of quantization bits, $\alpha$, with limited and perfect backhaul link.}
	\label{perf_imperfc_MN280_K40_ortho_vsalpha}
\end{figure}
\begin{figure}[t!]
	\center
	\includegraphics[width=80mm]{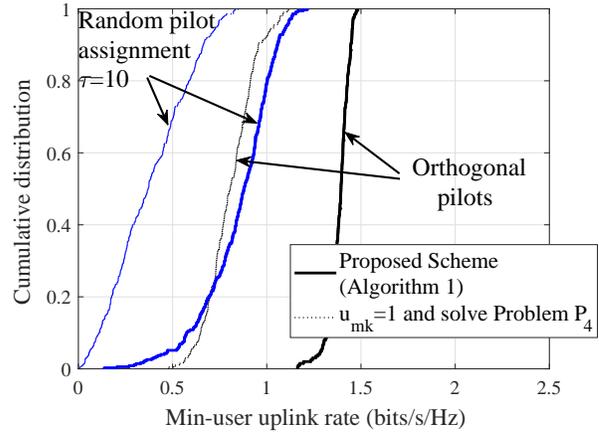}
	\vspace{-.09in}
	\caption{The cumulative distribution of the per-user uplink rate, for $M=100$, $N=2$, $K=40$, $\alpha=5$ and $D = 1$ km.}
	\label{compare_propo_100_2_40}
\end{figure}
\subsubsection{Effect of the Capacity of Backhaul Links}In Fig. \ref{perf_imperfc_MN280_K40_ortho_vsalpha}, a cell-free Massive MIMO system is considered with total number of service antennas $MN=280$, i.e. $(N=2,M=140)$, $(N=4,M=70)$ and $(N=10,M=28)$, $K=40$, orthogonal pilot sequences and $D=1$ km. As Fig. \ref{perf_imperfc_MN280_K40_ortho_vsalpha} shows, for the case of $(N=2,M=140)$ to achieve the performance of perfect backhaul links, we need to set $\alpha\ge 14$, for the cases of  $(N=4,M=70)$ and $(N=10,M=28)$, we need to set $\alpha\ge 13$ and $\alpha\ge 12$, respectively.
\subsubsection{Performance of the Proposed User Max-Min SINR Algorithm} Fig. \ref{compare_propo_100_2_40} presents the cumulative distribution
of the achievable uplink rates for the proposed Algorithm \ref{al1} and the scheme without considering the coefficients ${\bf{u}}_k$s similar to \cite{marzetta_free16} (we set $u_{mk}=1$, $\forall m,k$ and solve Problem $P_4$) with random pilot assignment with length $\tau=20$ and orthogonal pilot sequences. As seen in Fig. \ref{compare_propo_100_2_40}, the median of the cumulative distribution function (CDF) of the minimum uplink rate of the users is increased by $2.3$ times and $2.05$ times respectively with random and orthogonal pilots compared to the scheme in \cite{marzetta_free16}. Moreover, Fig. \ref{compare_propo_100_2_40} demonstrates that the rate of the proposed max-min SINR approach is more concentrated around the median value. 
%\begin{figure}[t!]
%\center
%\includegraphics[width=73mm]{con_100_40_ortho_t20.eps}
%\vspace{-.12in}
%\caption{The convergence of the proposed max-min SINR approach (Algorithm \ref{al1}) for $M=70$, $N=4$, $K=40$, $\tau=30$, $\alpha=5$ and $D=1$ km.}
%\label{con_100_40_ortho_t20}
%\end{figure}
\begin{figure}[t!]
\center
\includegraphics[width=80mm]{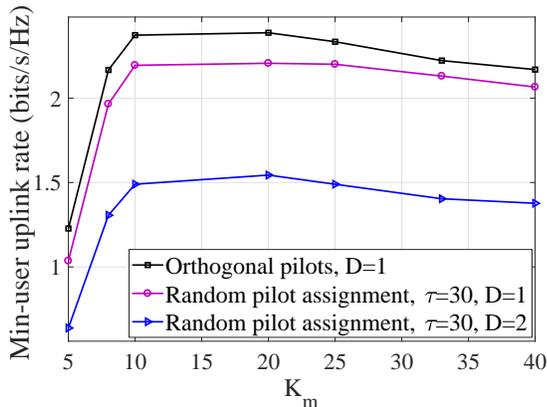}
\vspace{-.15in}
\caption{Average per-user uplink rate versus the total number of active users for each AP with  $M=100$, $N=2$, $K=40$ and $\log_2 Q_m\times K_{m}=200$.}
\label{assign_M100_K40}
\end{figure}
\begin{figure}[t!]
	\center
	\includegraphics[width=80mm]{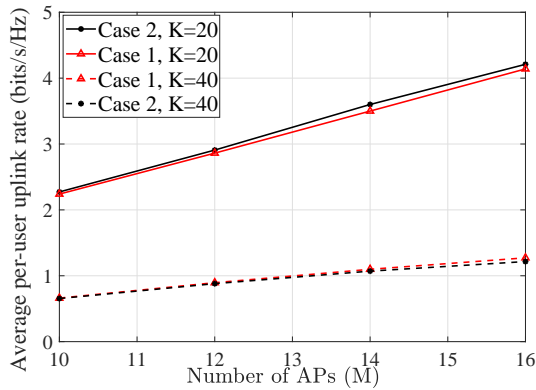}
	\caption{The average per-user uplink rate for cases 1 and 2, with ($N=4$, $K=20$, $\tau=20$, , $\alpha_1=9$, $\alpha_2=2$, $\omega_g=3$, $\omega_y=80$, $\omega_z=3$), and ($N=20$, $K=40$, $\tau=40$, $\alpha_1=8$, $\alpha_2=5$, $\omega_g=3.5$, $\omega_y=70$, $\omega_z=3$) with $D=1$ km and $\tau_c=200$.}
	\label{new2_k20_k40_hien_powercontrol}
\end{figure}
%\subsubsection{Convergence}Fig. \ref{con_100_40_ortho_t20} investigates the convergence of the proposed Algorithm {\ref{al1}} with $M=100$ APs and $K=40$ users and random pilot sequences with length $\tau=20$ and orthogonal pilot assignment. The figure confirms that the proposed algorithm converges in 2 iterations.
\subsubsection{Performance of the Proposed User Assignment Scheme} In Fig. \ref{assign_M100_K40}, the average per-user uplink rate is presented with $M=100$, $N=2$, $K=40$, orthogonal pilot sequences and random pilot assignment and with $D=1$ km and $D=2$ km versus the total number of active users per AP. Here, we used inequality (\ref{qk}) and set $\log_2 Q_m\times K_{m}=200$ for all curves in Fig. \ref{assign_M100_K40}. The optimum value of $K_m$, ($K_m^{\text{opt}}$), depends on the system parameters and as Fig. \ref{assign_M100_K40} shows for the case of $M=100$, $N=2$, $K=40$, the optimum value is achieved by $K_m^{\text{opt}}=20$. As a result, the proposed user assignment scheme can improve the performance of cell-free Massive MIMO systems with limited backhaul capacity. For instance, using  the proposed user assignment scheme for the case of $\tau=30$ and $D=2$ in Fig. \ref{assign_M100_K40}, one can achieve min-user uplink rate of $1.55$ $\text{bits}/\text{s}/\text{Hz}$ by setting $K_m^{\text{opt}}=20$, instead of quantizing the signals of all $K=40$ users and achieving min-user uplink rate of $1.37$ $\text{bits}/\text{s}/\text{Hz}$, which indicates $15\%$ improvement in the performance of cell-free Massive MIMO systems with limited backhaul capacity.
\subsubsection{Performance of Different Cases of Uplink Transmission}Fig. \ref{new2_k20_k40_hien_powercontrol} presents the average min-user uplink rate, where the per-user uplink rate is obtained by solving Problem $P_4$, given by (\ref{p4}) for Cases 1 and 2. In addition, for $K=20$, we set $\alpha_1=9$ and $\alpha_2=2$, for Case 1 and Case 2, respectively. The values of $\alpha_1=9$ and $\alpha_2=2$ correspond to a total number of 14,400 bits for each AP during each coherence time (or frame).
In addition, similar to \cite{Oppenheimsignal} we use a uniform quantzier with fixed stepsize. Simulation results show ($\omega_g=3$, $\omega_y=80$, $\omega_z=3$) are the optimal values for the case of $N=4$ and $K =20$, and ($\omega_g=3.5$, $\omega_y=70$, $\omega_z=3$) are the optimal values for the case of $N=20$ and $K=40$. As Fig \ref{new2_k20_k40_hien_powercontrol} shows the performance of Case 2 is better than Case 1 for $K=20$. Next, the performance of the cell-free Massive MIMO system is evaluated for a system with $K=40$ in which each AP is equipped with $N=20$ antennas.
Fig. \ref{new2_k20_k40_hien_powercontrol} shows the average rate of the cell-free Massive MIMO system, where for Case 1 and Case 2, we set $\alpha_1=3$ and $\alpha_2=8$, respectively which leads to a total number of 64,000 backhaul bits per AP per frame. Fig. \ref{new2_k20_k40_hien_powercontrol} shows that the performances of Case 1 and Case 2 depend on the values of $N$, $K$ and $\tau_f$. As in case 1, the CPU knows the quantized channel estimates, other signal processing techniques (e.g. zero-forcing processing) can be implemented to improve the system performance and can be considered in future work.
\section{Conclusions}
We have studied the uplink max-min SINR problem in cell-free Massive MIMO systems with the realistic assumption of limited capacity backhaul links, and have proposed an optimal solution to maximize the minimum uplink user rate. The numerical results confirmed that the proposed max-min SINR algorithm can increase 
the median of the CDF of the minimum uplink rate of the users by more than two times, compared to existing algorithms. We finally showed that further improvement ($15 \%$) in minimum rate of the users is achieved by the proposed user assignment algorithm.
\section*{Appendix A: Proof of Theorem \ref{theorem_up_quan_u}}
The desired signal for the user $k$ is given by
\begin{small}
\begin{IEEEeqnarray}{rCl}
\!\!\!\text{DS}_k\!=\!\sqrt{\rho}\mathbb{E}\!\left\{\!\sum_{m=1}^{M}\!u_{mk}\hat{\textbf{g}}_{mk}^H\textbf{g}_{mk}\!\sqrt{q_k}\right\}\!
=N\sqrt{\rho q_k}\!\sum_{m=1}^{M}\!u_{mk}\!\gamma_{mk}.
\label{dsk_vector}
\end{IEEEeqnarray}
\end{small}
The term $\mathbb{E}\{\left | \text{BU}_k\right |^2\}$ can be obtained as
\begin{IEEEeqnarray}{rCl}
\small
&&\mathbb{E} \left\{ \left | \text{BU}_k \right | ^2\right\} = \rho\mathbb{E}  \Biggl
\{ \Biggl| \sum_{m=1}^Mu_{mk}\hat{\textbf{g}}_{mk}^H{\textbf{g}}_{mk}\sqrt{q_k}\\
 &\!\!\!\!\!-&\!\!\rho \mathbb{E}\Big\{\sum_{m=1}^Mu_{mk}\hat{\textbf{g}}_{mk}^H{\textbf{g}}_{mk}\sqrt{q_k}\Big\}\Biggr|^2 \Biggr \} =\rho N\sum_{m=1}^Mq_ku_{mk}^2\gamma_{mk}\beta_{mk},\nonumber
\label{ebuk}
\end{IEEEeqnarray}
where the last equality comes from the analysis in \cite{marzetta_free16}, and using $\gamma_{mk}=\sqrt{\tau p_p}\beta_{mk}c_{mk}$.
The term $\mathbb{E}\{\left | \text{IUI}_{k k^\prime}\right |^2\}$ is obtained as 
%\vspace{-.45cm}
\begin{IEEEeqnarray}{rCl}
\small
\mathbb{E}& \{| &\text{IUI}_{k k^\prime} |^2 \}= \rho  \underbrace{q_{k^\prime} \mathbb{E}\left \{\left |\sum_{m=1}^Mc_{mk}u_{mk}\textbf{g}_{mk^\prime}^H\tilde{\bf{w}}_{mk}  \right |^2\right\}}_{A}
\nonumber\\
 &\!\!\!\!\!\!\!\!\!\!\!\!\!\!\!+&\!\!\!\!\!\!\!\rho \underbrace{\!\tau p_p\mathbb{E}\! \left \{ \!q_{k^\prime}\!\left | \! \sum_{m=1}^M\!c_{mk}u_{mk}\!\Big(\sum_{i=1}^{K}\textbf{g}_{mi}\!\pmb{\phi}_k^H
 \pmb{\phi}_i\!\Big)^H{\textbf{g}}_{mk^\prime}\!\right |^2 \right \} }_{B},
 \label{eiui}
\end{IEEEeqnarray}
%\vspace{-.05cm}
where the third equality in (\ref{eiui}) is due to the fact that for two independent random variables $X$ and $Y$ and $\mathbb{E}\{X\}=0$, we have $\mathbb{E}\{\left | X+Y \right |^2\}=\mathbb{E}\{\left | X \right |^2\}+\mathbb{E}\{\left | Y \right |^2\}$ \cite{marzetta_free16}.
Since $\tilde{\bf{w}}_{mk}=\pmb{\phi}_k^H\bf{W}_{p,m}$ is independent from the term $g_{mk^\prime}$ similar to \cite[Appendix A]{marzetta_free16}, the term $A$ in (\ref{eiui}) immediately is given by
$
A = N q_{k^\prime} \sum_{m=1}^Mc_{mk}^2u_{mk}^2\beta_{mk^\prime}.
$
The term $B$ in (\ref{eiui}) can be obtained as
%\vspace{-.05in}
\begin{IEEEeqnarray}{rCl}
 \label{bb111} 
B &=&  \underbrace{\tau p_p q_{k^\prime}\mathbb{E}\left \{\left | \sum_{m=1}^M c_{mk}u_{mk} ||{\textbf{g}}_{mk^\prime}||^2\pmb{\phi}_k^H{\pmb{\phi}}_{k^\prime}\right |^2\right \}}_{C}\\
 &+&\underbrace{\tau p_p q_{k^\prime}\mathbb{E} \left\{\left | \sum_{m=1}^Mc_{mk}u_{mk} \Big(\sum_{i\ne k^\prime}^{K}\textbf{g}_{mi}\pmb{\phi}_k^H\pmb{\phi}_i\Big)^H {\textbf{g}}_{mk^\prime}\right |^2 \right\}}_{D}.\nonumber
\end{IEEEeqnarray}
The first term in (\ref{bb111}) is given by
\begin{IEEEeqnarray}{rCl}
\small
C &=&N \tau p_p q_{k^\prime}\left |\pmb{\phi}_k^H{\pmb{\phi}}_{k^\prime}\right |^2\sum_{m=1}^Mc_{mk}^2u_{mk}^2\beta_{mk^\prime}\nonumber\\
 &+&
 N^2q_{k^\prime}\left |\pmb{\phi}_k^H{\pmb{\phi}}_{k^\prime}\right |^2\left(\sum_{m=1}^M u_{mk}\gamma_{mk}\dfrac{\beta_{mk^\prime}}{\beta_{mk}}\right)^2,
 \label{CC}
\end{IEEEeqnarray}
where the last equality is derived based on the fact $\gamma_{mk}=\sqrt{\tau p_p}\beta_{mk}c_{mk}$. The second term in (\ref{bb111}) can be obtained as
\begin{IEEEeqnarray}{rCl}
\small
D 
 &=&\!N\!\sqrt{\tau p_p}q_{k^\prime}\!\sum_{m=1}^{M}\!u_{mk}^2c_{mk}\beta_{mk^\prime}\beta_{mk}\!-\!Nq_{k^\prime}\!\sum_{m=1}^{M}\!u_{mk}^2c_{mk}^2\beta_{mk^\prime}\nonumber\\
 &-&N\tau p_p q_{k^\prime}\sum_{m=1}^{M}u_{mk}^2c_{mk}^2\beta_{mk^\prime}^2\left| \pmb{\phi}_k^H{\pmb{\phi}}_{k^\prime}\right|^2.
 \label{d}
\end{IEEEeqnarray} 
Finally by substituting (\ref{CC}) and (\ref{d}) into (\ref{bb111}), and substituting (\ref{bb111}) into (\ref{eiui}), we obtain
\begin{IEEEeqnarray}{rCl}
 \label{euiu}
\mathbb{E}\{| \text{IUI}_{k k^\prime}|^2\} &=&N\rho q_{k^\prime}\left(\sum_{m=1}^{M}u_{mk}^2\beta_{mk^\prime}\gamma_{mk}\right)\\
 &+&N^2 \rho q_{k^\prime} \left|\pmb{\phi}_k^H{\pmb{\phi}}_{k^\prime}\right|^2 \left(\sum_{m=1}^{M}u_{mk} \gamma_{mk}\dfrac{\beta_{mk^\prime}}{\beta_{mk}}\right)^2.\nonumber
\end{IEEEeqnarray}
The total noise for the user $k$ is given by
\begin{IEEEeqnarray}{rCl}
\!\!\!\!\!\!\!\!\!\!\!\!\!\mathbb{E}\!\left\{\!\left|\!\text{TN}_k\!\right|^2\!\right\}\!=\!\mathbb{E}\!\left\{\left|\!\sum_{m=1}^{M}u_{mk}\hat{\textbf{g}}_{mk}^H\textbf{n}_m\right|^2\!\right\}\!=\!N\sum_{m=1}^{M}u_{mk}^2\gamma_{mk},\!
\label{tn}
\end{IEEEeqnarray}
where the last equality is due to the fact that the terms $\hat{\textbf{g}}_{mk}$ and $\textbf{n}_m$ are uncorrelated.
The power of quantization error for the $k$th user is obtained as
\begin{small}
\begin{IEEEeqnarray}{rCl}
\!\!\!\!\!\!\!\!\!\mathbb{E}\!\left\{\!\left|\text{TQE}_k\!\!\right|^2\!\right\}\!=\!\frac{\!N\omega_z^2\!}{3Q_m^2}\!\sum_{m=1}^M\!u_{mk}^2\!\!\left[\!\left(2\beta_{mk}\!-\!\gamma_{mk}\!\right)\!\rho\!\sum_{k^\prime=1}^{K}\!q_{k^\prime}\!\beta_{mk^\prime}\!+\!\!\gamma_{mk}\!\right]\!\!,\!
\label{e6}
\end{IEEEeqnarray}
\end{small}
where the proof uses
a similar way to determine the power of quantziation error in Section II and is omitted here
due to space limitations.
Finally, SINR of the $k$th user is obtained by (\ref{sinr1}), which completes the proof of Theorem 3. ~~~~~~~~~~~~~~~~~~~~~$\blacksquare$
\bibliographystyle{IEEEtran}
\bibliography{limited_conf_debbah_hien_cameraready} 

% Generated by IEEEtran.bst, version: 1.14 (2015/08/26)
\begin{thebibliography}{10}
\providecommand{\url}[1]{#1}
\csname url@samestyle\endcsname
\providecommand{\newblock}{\relax}
\providecommand{\bibinfo}[2]{#2}
\providecommand{\BIBentrySTDinterwordspacing}{\spaceskip=0pt\relax}
\providecommand{\BIBentryALTinterwordstretchfactor}{4}
\providecommand{\BIBentryALTinterwordspacing}{\spaceskip=\fontdimen2\font plus
\BIBentryALTinterwordstretchfactor\fontdimen3\font minus
  \fontdimen4\font\relax}
\providecommand{\BIBforeignlanguage}[2]{{%
\expandafter\ifx\csname l@#1\endcsname\relax
\typeout{** WARNING: IEEEtran.bst: No hyphenation pattern has been}%
\typeout{** loaded for the language `#1'. Using the pattern for}%
\typeout{** the default language instead.}%
\else
\language=\csname l@#1\endcsname
\fi
#2}}
\providecommand{\BIBdecl}{\relax}
\BIBdecl

\bibitem{5gdebbah}
A.~Zappone, L.~Sanguinetti, G.~Bacci, E.~A. Jorswieck, and M.~Debbah,
  ``Energy-efficient power control: a look at {5G} wireless technologies,''
  \emph{IEEE Trans. Signal Process.}, vol.~64, no.~7, pp. 1668--1683, Apr.
  2016.

\bibitem{multidebbah}
E.~Björnson, E.~A. Jorswieck, M.~Debbah, and B.~Ottersten, ``Multiobjective
  signal processing optimization: the way to balance conflicting metrics in
  {5G} systems,'' \emph{IEEE Signal Process. Mag.}, vol.~31, no.~6, pp. 14--23,
  Oct. 2014.

\bibitem{ourvtc18}
A.~G. Burr, M.~Bashar, and D.~Maryopi, ``Cooperative access networks: Optimum
  fronthaul quantization in distributed {Massive} {MIMO} and cloud {RAN},'' in
  \emph{Proc. IEEE VTC}, Jun. 2018, pp. 1--5.

\bibitem{ouricc1}
M.~Bashar, K.~Cumanan, A.~G. Burr, , M.~Debbah, and H.~Q. Ngo, ``Enhanced
  max-min {SINR} for uplink cell-free {Massive} {MIMO} systems,'' in
  \emph{Proc. IEEE ICC}, May 2018, pp. 1--6.

\bibitem{ouriet_mic}
M.~Bashar, A.~G. Burr, K.~Haneda, and K.~Cumanan, ``Robust user scheduling with
  {COST} 2100 channel model for {Massive} {MIMO} networks,'' \emph{Accepted to
  publish in IET microwave antenna and propagation}, Feb. 2018.

\bibitem{slock_fromMU}
Y.~Lejosne, M.~Bashar, D.~Slock, and Y.~Yuan-Wu, ``From {MU} {Massive} {MISO}
  to pathwise {MU} {Massive} {MIMO},'' in \emph{Proc. IEEE SPAWC}, Jun. 2014,
  pp. 16--20.

\bibitem{marzetta_free16}
H.~Q. Ngo, A.~Ashikhmin, H.~Yang, E.~G. Larsson, and T.~L. Marzetta,
  ``Cell-free {Massive} {MIMO} versus small cells,'' \emph{IEEE Trans. Wireless
  Commun.}, vol.~16, no.~3, pp. 1834--1850, Mar. 2017.

\bibitem{ourjournal2}
M.~Bashar, K.~Cumanan, A.~G. Burr, , H.~Q. Ngo, and M.~Debbah, ``Max-min {SINR}
  of cell-free {Massive} {MIMO} uplink with optimal uniform quantization,''
  \emph{Submitted to IEEE Trans. Commun.}

\bibitem{backmacro}
Z.~Gao, L.~Dai, D.~Mi, Z.~Wang, M.~A. Imran, and M.~Z. Shakir, ``{MmWave}
  {Massive}-{MIMO}-based wireless backhaul for the {5G} ultra-dense network,''
  \emph{IEEE Trans. Wireless Commun.}, vol.~22, no.~5, pp. 13--21, Oct. 2015.

\bibitem{bookematrix}
G.~Golub and C.~V. Loan, \emph{Matrix Computations}, 2nd~ed.\hskip 1em plus
  0.5em minus 0.4em\relax Baltimore, MD: The Johns Hopkins Univ. Press, 1996.

\bibitem{gpboyd}
S.~P. Boyd, S.~J. Kim, A.~Hassibi, and L.~Vandenbarghe, ``A tutorial on
  geometric programming,'' \emph{Optim. Eng.}, vol.~8, no.~1, pp. 67--128,
  2007.

\bibitem{Oppenheimsignal}
A.~V. Oppenheim, R.~W. Schafer, and J.~R. Buck, \emph{Discrete-time signal
  processing}.\hskip 1em plus 0.5em minus 0.4em\relax Prentice-hall Englewood
  Cliffs, 1989.

\bibitem{cuma_sinr_spl10}
K.~Cumanan, L.~Musavian, S.~Lambotharan, and A.~B. Gershman, ``{SINR} balancing
  technique for downlink beamforming in cognitive radio networks,'' \emph{IEEE
  Signal Process. Lett.}, vol.~17, no.~2, pp. 133--136, Feb. 2010.

\end{thebibliography}
\end{document}